\documentclass[aps,prx,reprint, amsmath,amssymb]{revtex4-2}

\usepackage{siunitx}
\usepackage[version=4]{mhchem}
\usepackage[utf8]{inputenc}
\usepackage{nicefrac}
\usepackage{graphicx}

\usepackage{epstopdf}

\begin{document}

\title{Opposite effects of the rotational and translational energy on the rates of ion-molecule reactions near \SI{0}{K}: the $\ce{D2+}+\ce{NH3}$ and $\ce{D2+}+\ce{ND3}$ reactions}
\author{Raphaël Hahn$^{1,2}$}
\author{David Schlander$^1$}
\author{Valentina Zhelyazkova$^{1,2}$}
\author{Frédéric Merkt$^{1,2,3}$}

\email[ Corresponding author: ]{frederic.merkt@phys.chem.ethz.ch}

\affiliation{$^1$ Department of Chemistry and Applied Biosciences, ETH Zurich, CH-8093 Zurich, Switzerland}
\affiliation{$^2$ Quantum Center, ETH Zurich, CH-8093 Zurich, Switzerland}
\affiliation{$^3$ Department of Physics, ETH Zurich, CH-8093 Zurich, Switzerland}

\date{\today}


\begin{abstract}
The ion-molecule reactions $\ce{D2+}+\ce{NH3}$ and $\ce{D2+}+\ce{ND3}$ are studied at low collision energies ($E_{\text{coll}}$ from zero to $\sim k_\textrm{B}\cdot \SI{50}{K}$), with the \ce{D2+} ions in the ground rovibrational state and for different rotational temperatures of the ammonia molecules, using the Rydberg-Stark merged-beam approach.
Two different rotational temperatures ($\sim\SI{15}{K}$ and $\sim\SI{40}{K}$), measured by (2+1) resonance-enhanced multiphoton-ionization spectroscopy, are obtained by using a seeded supersonic expansion in He and a pure ammonia expansion, respectively.  
The experimental data reveal a strong enhancement of the rate coefficients at the lowest collision energies caused by the charge-dipole interaction.
Calculations based on a rotationally adiabatic capture model accurately reproduce the observed kinetic-energy dependence of the rate coefficients. 
The rate coefficients increase with increasing rotational temperature of the ammonia molecules, which contradicts the expectation that rotational excitation should average the dipoles out. Moreover, these reactions exhibit a pronounced inverse kinetic isotope effect.
The difference is caused by nuclear-spin-statistical factors, and the smaller rotational constants and tunneling splittings in \ce{ND3}.
\end{abstract}

\maketitle

\section{Introduction}
Understanding how the rates of chemical reactions depend on the quantum states of the reactants and on the collision energy is one of the fundamental goals of chemical physics \cite{levine05a,dulieu17a}. Particularly strong effects are expected for ion-molecule reactions at low temperatures because long-range electrostatic interactions lead to strongly state-specific anisotropic potentials and to a large collision-energy dependence of the reaction rates \cite{ng92a,heazlewood21a}. 
Recent progress in the control of the motion and quantum states of molecular ions makes it possible to investigate ion-molecule reactions with unprecedented details \cite{chang_specific_2013,puri19a,markus20a,zhelyazkova20a,wild_tunnelling_2023}.
In the present article, we report on measurements of the fundamentally important reaction between state-selected molecular-hydrogen ions and ammonia near \SI{0}{K} which reveal, for the first time, remarkably strong effects of the collision energy, the rotational temperature and isotopic substitution.

Exothermic, barrier-free ion-molecule reactions are the dominant chemical processes in cold dilute environments and play a key role in astrophysics and plasma physics \cite{herbst73a,hartquist96a,larsson12a,smith11a,jimenezredondo11a}.
Most of the ion-molecule reaction rates (and their branching ratios) required for the modeling of the chemical composition of the interstellar medium (ISM) and planetary atmospheres have not been measured, or have been measured at room temperature \cite{carrasco_sensitivity_2008,westlake_role_2014}. 
These rates and branching ratios are usually considered constant over the \SI{3}{K} to \SI{200}{K} temperature range, an assumption that introduces uncertainties and errors in the ion and neutral concentrations predicted by global kinetic models \cite{carrasco_sensitivity_2008}. 
At low temperature, the attractive electrostatic long-range interactions between ions and molecules imply large capture rate coefficients that can exceed the Langevin capture rate coefficients when the molecules have permanent dipole and quadrupole moments (see Ref. \cite{heazlewood21a} and references therein).
The dependence of the ion-molecule reaction rates on the vibrational and/or rotational temperature of the neutral molecule
is mostly unknown but is expected to play an important role, particularly for regions of the ISM deviating from local thermodynamic equilibrium \cite{indriolo_constraining_2010}.

Despite their importance, few measurements of rate coefficients for ion-molecule reactions below \SI{100}{K} have been reported so far.
The lack of low collision-energy experimental data comes partly from the fact that even very weak stray fields accelerate the ions and significantly heat them up, i.e. an electric potential difference of \SI{1}{mV} imparts a kinetic energy of $k_\text{B}\cdot\SI{12}{K}$ to the ion. Space-charge-induced repulsion and heating also severely limit the ion densities, which can become incompatible with low temperature investigations. In addition, the common methods used to produce molecular ions such as photoionization or electron-impact ionization indeed usually lead to ion populations distributed over multiple vibrational and rotational states.

Experimental tools and methods developed to overcome these challenges and reach low-temperature conditions in the study of ion-molecule reactions include cold uniform supersonic flows \cite{marquette85a,potapov_uniform_2017,rowe_ion_2022}, with which temperatures as low as $\sim\SI{10}{K}$ can be reached; buffer-gas cooling in ion traps \cite{markus20a,kumar_low_2018,tran18a,venkataramanababu_enhancing_2023,wild_tunnelling_2023}, with which collision-energy down to $\sim 3\cdot k_\text{B} - 5\cdot k_\text{B}$ K are accessible; laser-cooled ions and Coulomb crystals \cite{willitsch08a,chang_specific_2013,krohn_isotope-specific_2021,okada_rotational_2022}; and superimposed traps of neutral atoms/molecules and ions \cite{puri19a,voute_charge_2023}, with which collision-energy well below $1\cdot k_\text{B}~K$ are achievable, however, at the expense of tunability and general applicability.

In the present work, we report on the investigation of the reactions
\begin{align}
    \ce{D2+}+\ce{NH3}&\longrightarrow \ce{D2}+\ce{NH3+},\\
    \ce{D2+}+\ce{NH3}&\longrightarrow \ce{H}+\ce{D}+\ce{NH2D+},
\end{align}
and  
\begin{align}
\ce{D2+}+\ce{ND3}&\longrightarrow \ce{D2}+\ce{ND3+},\\
\ce{D2+}+\ce{ND3}&\longrightarrow \ce{D}+\ce{D}+\ce{ND3+}
\end{align}
with state-selected \ce{D2+} $(v^+=0,N^+=0)$ ions at collision energies in the range between zero and $\sim k_\text{B}\cdot\SI{50}{K}$ using the Rydberg-Stark merged-beam approach \cite{allmendinger16a,zhelyazkova20a}. \ce{D2+} was studied instead of \ce{H2+} because of the lower terminal velocity of the \ce{D2} molecular beam, which provides easier access to zero relative velocity with the beam of ground-state molecules.
With this method, ion-molecule reactions are observed within the orbit of a highly excited Rydberg electron that shields the reaction systems from stray fields and affects neither their kinetics nor their outcome \cite{allmendinger16a,martins21a}. 
This technique exploits the large dipole moments of Rydberg-Stark states (up to $\sim\SI{3400}{D}$ at $n=30$) to deflect Rydberg-atom and -molecule beams and merge them with beams of neutral ground-state molecules.
The spectator role of the electron can be understood in the realm of the independent-particle model of collisions involving Rydberg atoms and molecules \cite{stebbings83a,pratt94a,wrede05a,matsuzawa10a}.

Three main scientific questions are at the focus of this investigation. First, we examine the dependence of the rate coefficients to the internal rotational excitation of the ammonia molecules, which we find to be positive.
Until very recently \cite{okada_rotational_2022}, only the effects of the rotational energy of the ion have been studied experimentally \cite{dressler_study_2006,malow_ionmolecule_2001,venkataramanababu_enhancing_2023} and found to be significant. It is, however, the rotational energy of the neutral molecules that is expected to lead to the strongest effects.
Second, we determine the dependence of the rate coefficients on the collision energy, which we find to be negative.
Finally, we clarify the origin of a surprisingly large inverse kinetic isotope effect observed in cold ion-molecule reactions involving \ce{NH3} and \ce{ND3} \cite{petralia20a,tsikritea_inverse_2021,ard_inconsistent_2022,petralia_reply_2022,toscano20a}.
The opposite effect of the internal rotational energy and the translational energy on the rate coefficients was unexpected at the outset, because charge-dipole interactions average out at high rotational excitation of the neutral molecule.
This apparent contradiction is interpreted using a rotationally adiabatic capture model.

\section{Theoretical modeling of the reaction rate coefficients}
\subsection{The Langevin model: shortcomings and extensions}

\begin{figure}
    \centering
    \includegraphics[width=\linewidth]{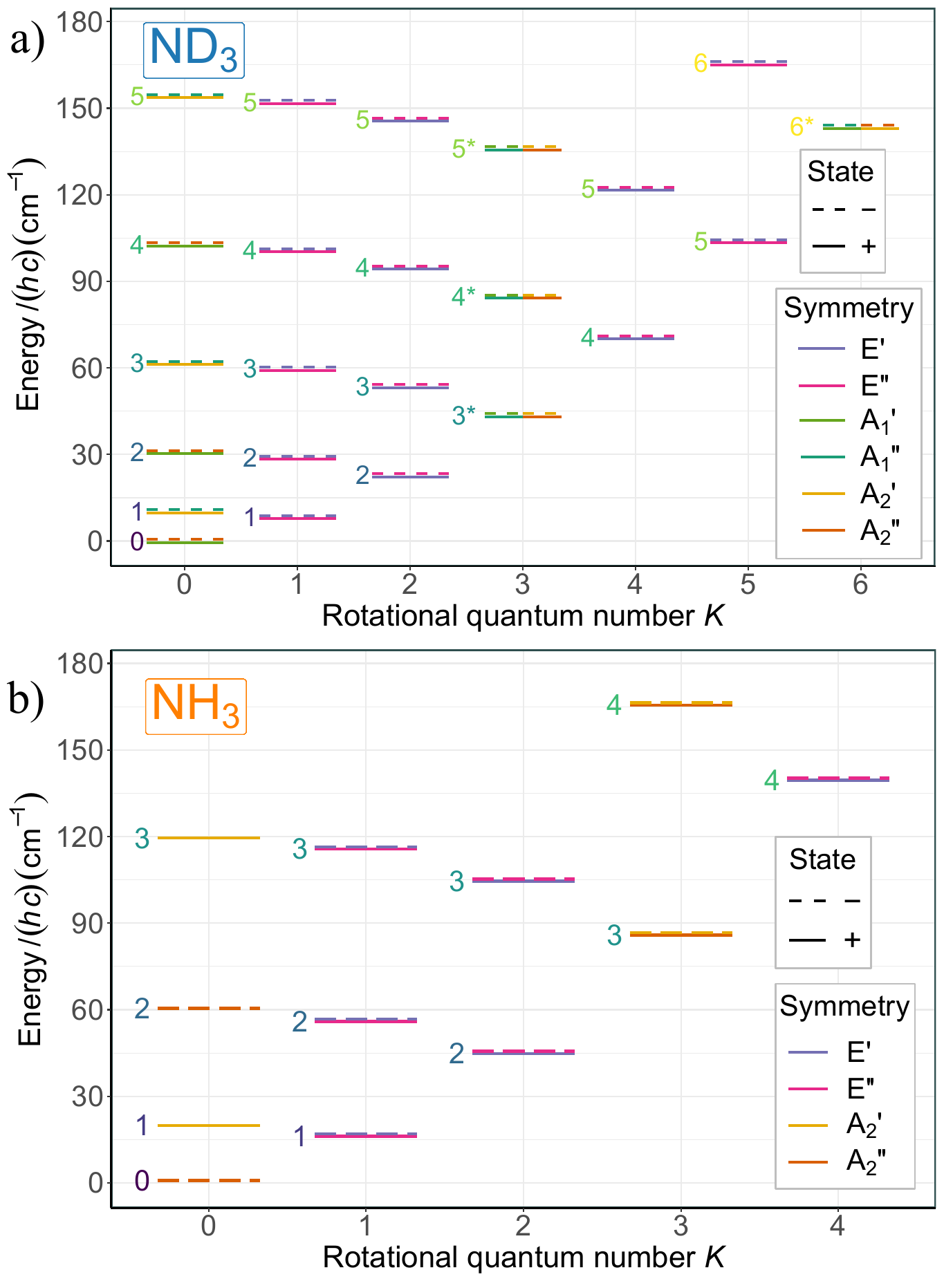}
    \caption{Rotational level structure of \ce{ND3} (a) and \ce{NH3} (b) in their $X \ce{^{1}A_{1}}'$ ground electronic state. The rovibronic symmetry of the $(J,K,p)$ states in the $D_{3h}$ molecular-symmetry group, and the value of $p$ and $J$ are indicated by the line color, the line type and by the number on the left of the lines, respectively. In a), asterisks indicate degenerate states of different symmetry. 
    The tunneling splittings in \ce{ND3} are expanded by a factor 20 for clarity.
    \label{fig:NH3andND3states}
    }
\end{figure}

The usual reference point for the evaluation of ion-molecule reaction rate coefficients is the Langevin model, which describes the interaction between the ion and the molecule as arising from the charge-induced-dipole interaction. 
The effective Langevin potential $V_{\textrm{eff,L}}(R)$ of the interaction of an ion and a neutral molecule at a distance $R$, including the centrifugal potential related to the collision angular-momentum $\vec{L}$, is given by \cite{merkt19a}
\begin{equation}
    V_{\textrm{eff,L}}\left(R\right)=-\frac{\alpha e^2}{32\pi^2\varepsilon_0^2 R^4}+\frac{\vec{L}^2}{2\mu R^2},
\end{equation}
where $\alpha$ is the polarizability of the neutral molecule, $e$ the elementary charge, $\varepsilon_0$ the vacuum permittivity and $\mu$ the reduced mass of the collision. 
The model assumes that every collision with collision energy $E_\textrm{coll}\geq V^{\text{max}}_{\textrm{eff,L}}\left(R\right)$ leads to a reaction, where $V^{\text{max}}_{\textrm{eff,L}}$ is the maximum of the $L$-dependent interaction potentials. This assumption results in reaction rate coefficients \cite{langevin05a}
\begin{equation}\label{eq:Langevin}
    k_{\textrm{L}}=\sqrt{\frac{\alpha e^2}{4\varepsilon_0\mu}}
\end{equation}
that are independent of $E_\textrm{coll}=\frac{1}{2}\mu v_{\textrm{rel}}^2$, and hence of the asymptotic relative velocity of the reactants $v_{\textrm{rel}}$. 
 The Langevin model, while useful as a reference point, does not accurately describe the reaction rates of ions with dipolar molecules. The model can be extended by adding a parameterized charge-dipole interaction as, e.g., in the average-dipole-orientation (ADO) \cite{su75a} or in the Su-Chesnavich \cite{su82a} approaches.
 However, these classical, semi-empirical models fail when considering cold reactants with only few occupied quantum states \cite{clary85a,clary85b,troe87a}, as demonstrated recently in the reactions of \ce{He+} with \ce{CH3F} \cite{zhelyazkova20a}, and \ce{NO} \cite{zhelyazkova23a}.

\begin{figure*}
    \centering
    \includegraphics[width=\textwidth]{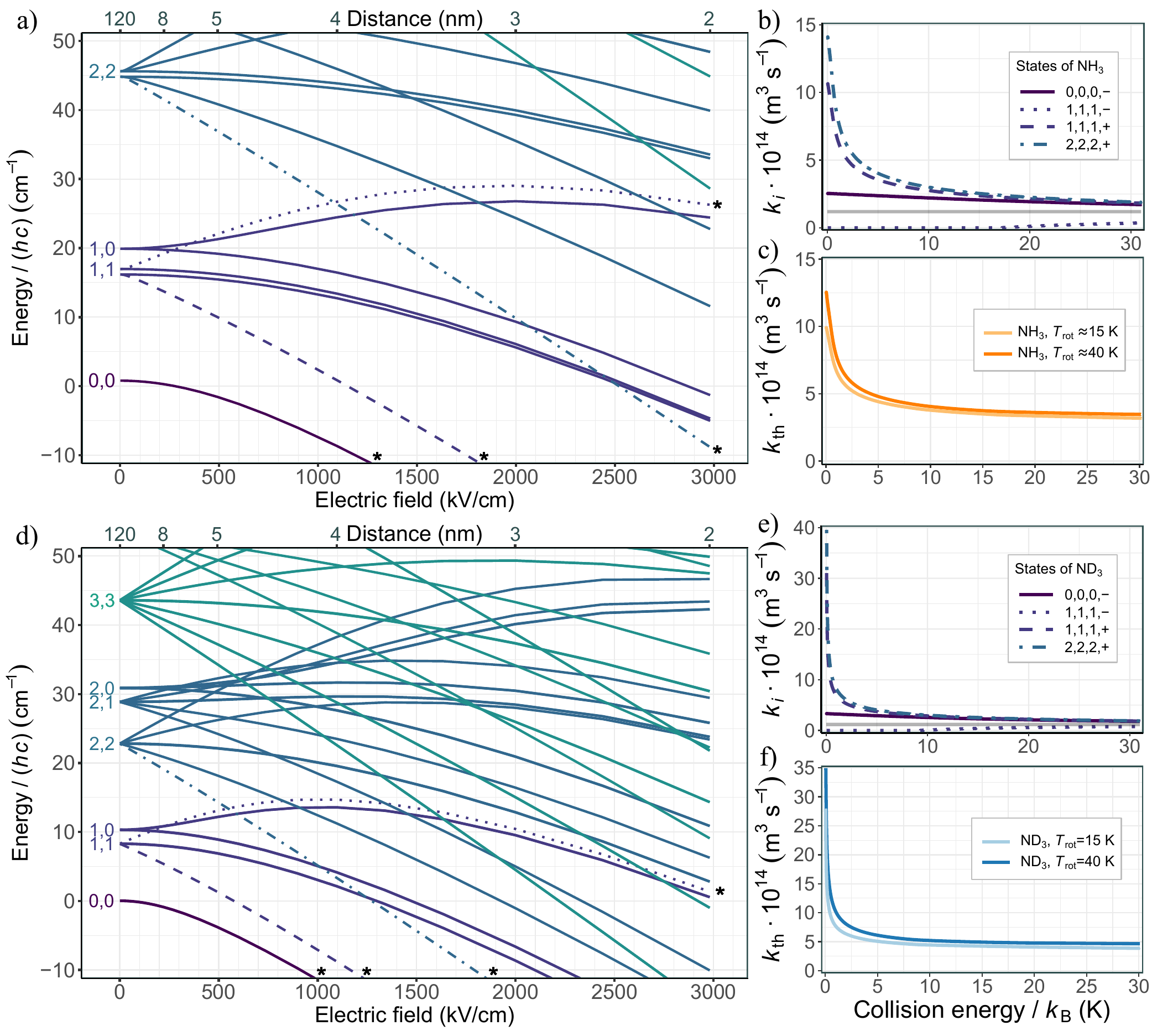}
    \caption{Rotationally adiabatic capture model for \ce{NH3} (a-c) and \ce{ND3} (d-f). a) Stark map of the rotational levels of \ce{NH3} in the electric field of a point-like ion at a given distance (top axis). The rotational states are labeled on the left by their quantum numbers $(J,K)$. 
    b) State-specific reaction rate coefficients of the four states labeled with an asterisk in a): $J=0,K=0,|M|=0,p=-$ (solid, dark purple), $J=1,K=1,|M|=1,p=+$ (dashed, indigo), $J=1,K=1,|M|=1,p=-$ (dotted, indigo) and $J=2,K=2,|M|=2,p=+$ (dash-dotted, teal). The Langevin rate coefficient is indicated by the grey horizontal line.
    c) Averaged reaction rate coefficients $k_{\text{th}}(E_{\text{coll}},T_{\textrm{rot}})$ obtained using the calculated state-specific rate coefficients and the occupation probabilities of the rotational states for an expansion of pure \ce{NH3} ($T_{\textrm{rot}}\approx \SI{40}{K}$) (dark orange) and a seeded expansion (5$\%$) in He ($T_{\textrm{rot}}\approx \SI{15}{K}$) (light orange).
    d)-f) Same as a)-c) but for \ce{ND3}.
    }
    \label{fig:theory}
\end{figure*}

\subsection{Adiabatic capture model}\label{sec:model}

To describe theoretically the reaction rates of ion-molecule reactions involving dipolar and quadrupolar molecules, we implement a rotationally adiabatic capture model inspired by the pioneering works of Clary \cite{clary85a,clary85b} and Troe \cite{troe87a} and described in detail in Refs. \cite{zhelyazkova20a,zhelyazkova21a}.  
The rotational-energy-level structures in the $X \ce{^{1}A_{1}}'$ ground electronic state of \ce{ND3} and \ce{NH3} are depicted in Panels  a) and b) of Fig. \ref{fig:NH3andND3states}, respectively.
\ce{NH3} and \ce{ND3} are symmetric-top molecules and their rotational states are labeled $(J,K,M,p)$ where $(J,K,M)$ are, respectively, the rotational-angular-momentum quantum number and the quantum numbers associated with the projections of the rotational-angular-momentum vector on the principal symmetry axis of the molecular reference frame and the $z$-axis of the laboratory frame, typically chosen along the direction of the electric field.
Their equilibrium geometry is pyramidal with a $C_{3v}$ symmetry, leading to two potential wells separated by a barrier along the umbrella-inversion vibrational mode, and quantum-mechanical tunneling between the two wells. 
The tunneling leads to a splitting of every $(J,K)$ state into a doublet. The lower and upper states of the doublet are the symmetric ($p=+$) and antisymmetric ($p=-$) superpositions of wavefunctions localized in the two wells.
Rotational-energy-level diagrams with full $(J,K,M,p)$ labels and nuclear-spin-statistical factors can be found in Refs. \cite{wichmann20a,zhelyazkova21a}. 
\ce{ND3} has a smaller rotational constant and thus a higher density of states than \ce{NH3}.
Moreover, the $(0,0,0,+)$, $(1,0,M,-)$ and $(2,0,M,+)$ states are not populated in \ce{NH3} because of restrictions imposed by the Pauli principle.

To calculate state-specific rate coefficients, we implement the rotationally adiabatic capture model in the same way as for the \ce{He+}+\ce{NH3} (\ce{ND3}) reactions in Ref. \cite{zhelyazkova21a}, where more details can be found. 
State-specific potentials $V_i(R)$ are obtained by adding the Stark shifts $\Delta E^{\text{Stark}}_{J,K,M,p} (R)=\Delta E^{\text{Stark}}_{i}(R)$ of the rotational levels in the field of the ion to the Langevin potential $ V_{\textrm{eff,L}}\left(R\right)$
\begin{equation}
V_i(R)=V_{\textrm{eff,L}}\left(R\right)+\Delta E^{\text{Stark}}_{i}(R).
\end{equation}
For each of these potentials 
and for a given collision energy $E_{\text{coll}}$, we calculate the maximum collision angular-momentum $L_{i,\text{max}}$ that leads to the capture of the neutral molecule in the field of the ion. The state-specific and collision-energy-dependent capture rates $k_i(E_{\text{coll}})$ are then given by \cite{zhelyazkova21a}
\begin{equation}
    k_i(E_{\text{coll}})=\frac{\pi L_{i,\text{max}} ^2}{\sqrt{2 \mu^3 E_{\text{coll}}}}.
\end{equation}
The procedure is illustrated in Fig. \ref{fig:theory}, which depicts the Stark shifts of the lowest rotational levels of \ce{NH3} and \ce{ND3} in panels a) and d), respectively. Panels b) and e) show the calculated state-specific capture rate coefficients for the four states $(0,0,0,-)$ (solid, dark purple,, $(1,1,1,+)$ (indigo, dashed), $(1,1,1,-)$ (indigo, dotted), $(2,2,2,+)$ (teal, dash-dotted) of \ce{NH3} and \ce{ND3}, respectively. 
States with negative Stark shifts show an increase of the capture rate coefficients as the collision energy decreases below \SI{20}{K}, whereas the capture rate coefficients of the states with positive Stark shifts typically vanish below \SI{10}{K}.
At high collision energies, the state-specific rate coefficients tend towards the Langevin rate coefficient (in grey). 
The theoretical capture rate coefficient
\begin{equation}\label{eq:kth}
    k_{\text{th}}(E_{\text{coll}},T_{\textrm{rot}})=\sum_i P_{i,T_{\textrm{rot}}} k_i(E_{\text{coll}})
\end{equation}
for a given rotational temperature $T_{\textrm{rot}}$ of the ammonia sample is obtained by multiplying these state-specific rates with the corresponding rotational-state occupation probabilities $P_{i,T_{\textrm{rot}}}$. In the present work, we measure these probabilities using (2+1) resonance-enhanced multiphoton ionization (REMPI) spectroscopy. The values of $k_{\text{th}}$ for experimentally determined rotational-state occupation probabilities of \ce{NH3} and \ce{ND3} (corresponding to rotational temperatures of \SI{15}{K} and \SI{40}{K}) are displayed in Panels c) and f) of Fig. \ref{fig:theory}, respectively.

\section{Experimental methods}\label{sec:methods}
\subsection{Merged-beam experimental set-up}\label{sec:exp}
\begin{figure*}
    \centering
    \includegraphics[width=\textwidth]{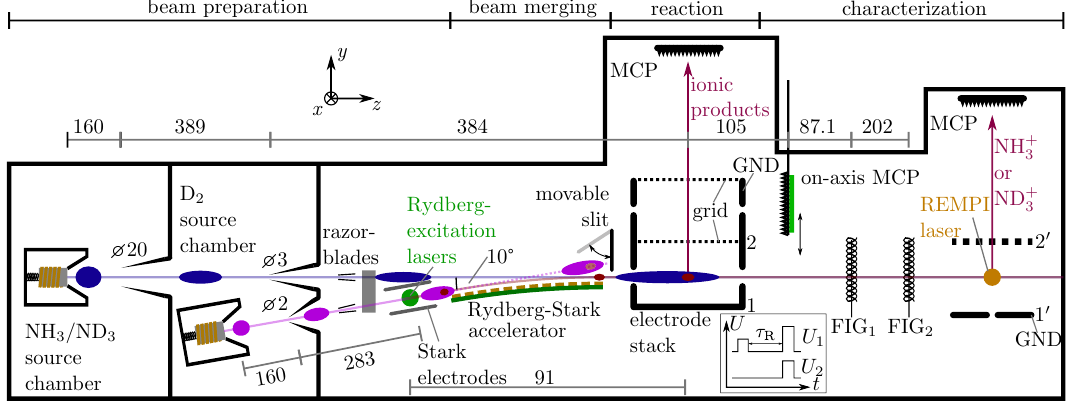}
    \caption{Schematic view of the experimental setup (not to scale). MCP: microchannel-plate detector, FG: fast-ionization gauge, GND: ground potential. 
A beam of Rydberg \ce{D2} molecules is formed by photoexcitation in a supersonic expansion of pure \ce{D2} between two electrodes. 
It is merged with a beam of either pure \ce{NH3} (\ce{ND3}) or \ce{NH3} (\ce{ND3}) seeded in He with a 5:95 ratio using a Rydberg-Stark accelerator. 
Cationic products of the reaction are extracted towards an MCP detector by applying pulsed electric potentials $U_1$ and $U_2$ (see inset) on electrodes 1 and 2, respectively, and their masses are deduced from their time-of-flight.
A movable imaging MCP allows the characterization of the accelerated $\ce{D2}(n)$ molecules. 
FG$_1$ and FG$_2$ are used to measure the velocity distribution and density of the ammonia molecules.
A permanent electric field is applied between electrodes $1^\prime$ and $2^\prime$ to extract photo-ionization products (\ce{NH3+} or \ce{ND3+}) towards an MCP detector. All distances are in mm. 
}
    \label{fig:setup}
\end{figure*}

The experimental setup and procedure are described in Ref. \cite{hoeveler21a}, and only the main aspects and the relevant modifications are presented in this section.
A schematic view of the setup is shown in Fig. \ref{fig:setup}.
The reactions are studied using supersonic molecular beams produced by two home-made pulsed valves producing short ($\approx$ $20~\mu$s) pulses of gas at a repetition rate of \SI{25}{Hz}. 
\ce{NH3} and \ce{ND3} are used either pure or in a (5:95) mixture with He. 
Helium was chosen as the carrier gas to inhibit the formation of ammonia clusters \cite{shinohara_twophotonionization_1983} and to increase the mean velocity of the beam.
Two skimmers (with diameters of \SI{20}{mm} and \SI{3}{mm}) and two pairs of razor blades constrain the size and the transverse velocity of the ammonia beam.
The second beam (hereafter referred to as the $\ce{D2}(n)$ beam) is formed from a skimmed beam of \ce{D2} molecules via resonant three-photon excitation to Rydberg-Stark states of principal quantum number $n=29$ belonging to the Rydberg series converging to the \ce{D2+}$(v^+=0,N^+=0)$ ionization threshold.
The photoexcitation takes place between two electrodes used to generate an electric field of \SI{10}{V/cm}.
This electric field mixes Rydberg states of different values of the orbital angular momentum quantum number $l$, which results in Rydberg-Stark states with large dipole moments that are sensitive to electric-field gradients.
The photoexcitation is carried out near the surface of an on-chip Rydberg-Stark deflector \cite{allmendinger16a}, with which time-dependent electric fields are applied to trap, deflect and accelerate the Rydberg molecules. 
We use this Rydberg-Stark deflector to merge the $\ce{D2}(n)$ and the ammonia beams, which initially propagate along axes separated by $10^{\circ}$, and to set the velocity of the $\ce{D2}(n)$ molecules.
For a $\ce{D2}(n)$ beam traveling initially at \SI{1500}{m/s}, the final velocity $v_{\text{f}}$ can be adjusted in the range between $\SI{1000}{m/s}$ and $\SI{2100}{m/s}$, corresponding to collision energies ranging from $E_{\text{coll}}/k_\text{B}\approx 0$ K to $E_{\text{coll}}/k_\text{B}\approx50$ K (see Fig. \ref{fig:Ecoll} below).

The merged beams enter the reaction chamber, comprising a time-of-flight mass spectrometer in a Wiley-McLaren configuration. An adjustable slit (see Fig. \ref{fig:setup}) blocks the untrapped Rydberg molecules from entering the reaction chamber.
The positively charged reaction products and \ce{D2+} ions produced by field-ionization of the $\ce{D2}(n)$ molecules are extracted towards a microchannel-plate (MCP) detector by applying two precisely timed electric-field pulses of different amplitudes across the reaction volume.
The first, weaker pulse (pre-pulse), removes all ions from the reaction region and defines the beginning of the reaction-observation temporal window.
The second electric-field pulse is applied when the cloud of Rydberg molecules reaches the center of the reaction region and extracts the product ions. The field-free interval between the two pulses represents the reaction-observation time $\tau_\text{R}$, and is kept constant at 14~$\mu$s for all measurements. The density of ammonia molecules is much larger than that of $\ce{D2}(n)$. Moreover, the molecule densities are such that the reaction probability of $\ce{D2}(n)$ is less than $1\%$. Consequently, the reaction rates are well described by pseudo-zero-order kinetics.
In preliminary experiments, we verified that the integrated product-ion signals are proportional to $\tau_\text{R}$, as expected for pseudo-zero-order kinetics.

For each selected collision energy, the ion signals are averaged over typically 5000 experimental cycles by a fast digital oscilloscope. 
To remove the contributions from reactions of $\ce{D2}(n)$ with the background gas, time-of-flight spectra are also recorded under conditions where the pulses of $\ce{D2}(n)$ and ammonia molecules do not overlap temporally in the reaction region. These background time-of-flight spectra are then subtracted from the recorded traces.
The velocity and temporal distributions of the $\ce{D2}(n)$ and ammonia beams are characterized using (see Fig. \ref{fig:setup}):
\begin{itemize}
    \item two fast-ionization gauges (FG$_1$ and FG$_2$ in Fig. \ref{fig:setup}) to measure the temporal profile of the neutral beam at two different positions. From this measurement, the velocity and spatial distributions of the ammonia beam can be inferred (see Section \ref{sec:neutr}).
    \item a movable imaging MCP detector (on-axis MCP) that can be slid into the beam to record the size of the $\ce{D2}(n)$ molecular cloud and its temporal profile. In this way, the transverse and the longitudinal velocity distributions of the $\ce{D2}(n)$ beam can be precisely determined (see Section \ref{sec:Ryd}).
    \item a REMPI chamber equipped with a pair of electrodes and located beyond the reaction chamber, used to record (2+1) REMPI spectra of the ammonia samples and determine their rotational temperature (see Section \ref{sec:REMPI}). 
\end{itemize}

\subsection{Beam characterization and determination of the collision energies}\label{sec:images}
The collision energy $E_{\text{coll}}$ in our merged-beam experiment is given by
\begin{equation}\label{eq:Ecoll}
    E_{\text{coll}}=\frac{1}{2}\mu ||\vec{v}_{\text{rel}}||^2=\frac{1}{2}\mu v_{\text{rel}}^2,
\end{equation}
with 
\begin{equation}\label{eq:vrel1}
|v_{\text{rel}}|=\sqrt{(v_{\text{Ry},x}-v_{\text{n},x})^2+(v_{\text{Ry},y}-v_{\text{n},y})^2+(v_{\text{Ry},z}-v_{\text{n},z})^2}.
\end{equation}
In Eq. \eqref{eq:vrel1}, $v_{\text{n},x},v_{\text{n},y},v_{\text{n},z}$ and $v_{\text{Ry},x},v_{\text{Ry},y},v_{\text{Ry},z}$ are the components of the velocity vectors of the ammonia and $\ce{D2}(n)$ molecules, respectively (see Fig. \ref{fig:setup} for the definition of the $x,y,z$ directions).
To reliably determine the collision-energy dependence of the reaction rate coefficients, it is essential
to determine the collision-energy distribution $\rho(E_{\text{coll}};v_{\text{f}})$ for each selected value of the mean final longitudinal velocity $v_{\text{f}}$ of the $\ce{D2}(n)$ cloud.

The determination of $\rho(E_{\text{coll}};v_{\text{f}})$ thus requires knowledge of the three-dimensional relative-velocity distributions $\rho(\vec{v}_{\text{rel}};v_{\text{f}})$ of the ammonia and $\ce{D2}(n)$ molecules in the reaction volume for each $v_{\text{f}}$.
The quantity $\rho(\vec{v}_{\text{rel}};v_{\text{f}})$ is derived from independent measurements of the three-dimensional velocity of the ammonia and $\ce{D2}(n)$ beams.
\subsubsection{Characterization of the ammonia beam} \label{sec:neutr}
\begin{figure}
    \centering
    \includegraphics[width=\linewidth]{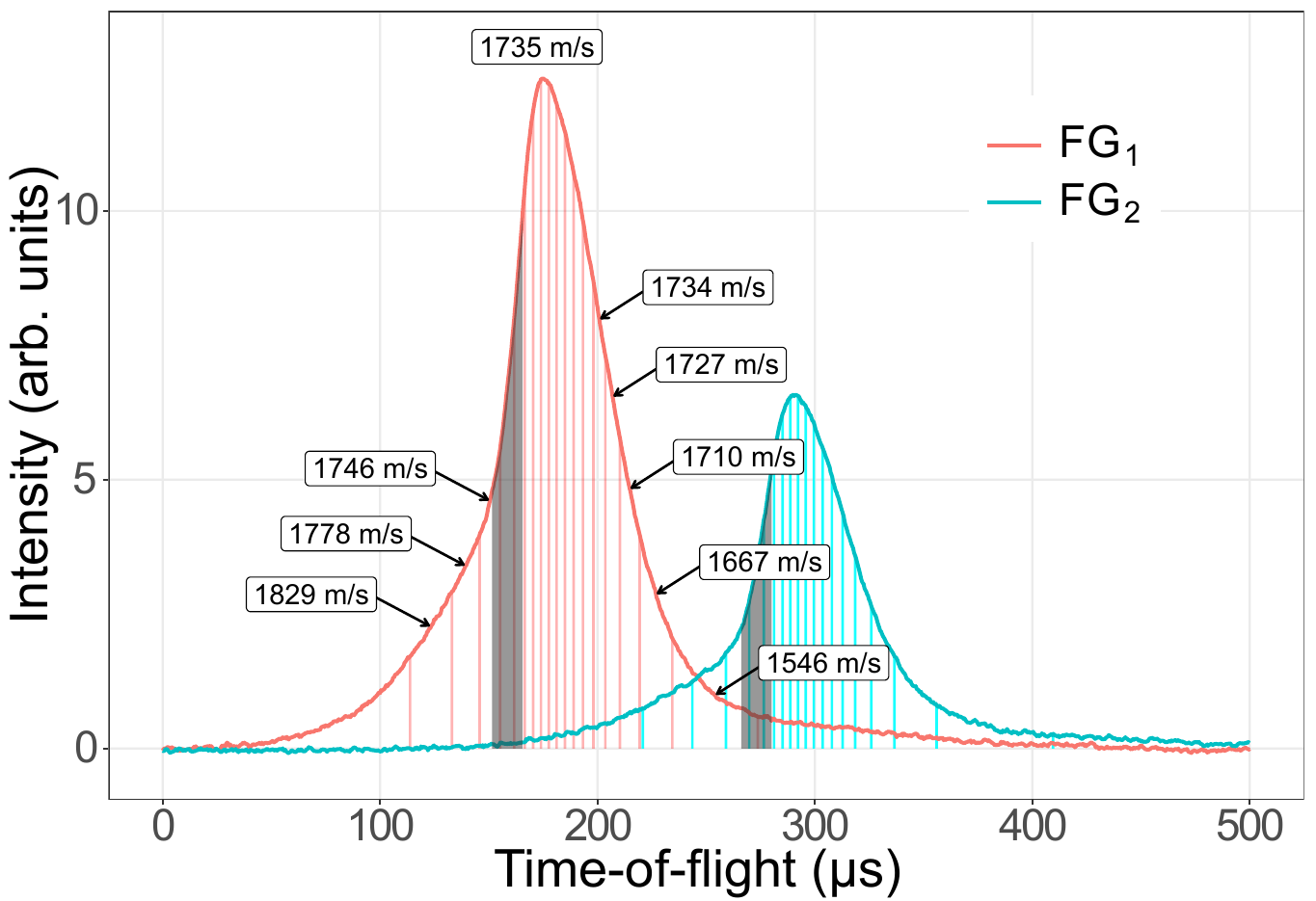}
    \caption{Determination of the velocity $\overline{v_{\text{n},z}}$ of the \ce{NH3} beam seeded in He from the time-of-flight distributions recorded at the two FGs.
    The time-of-flight profiles are divided into 20 bins with equal areas under the curve (indicated with red and blue vertical lines for the first and second FG, respectively). 
    The area shaded in grey on each FG profile corresponds to the molecules overlapping with the $\ce{D2}(n)$ cloud during the 14-$\mu$s-long reaction-observation temporal window. This area is used for normalization of the product-ion signal.
    The origin of the time-of-flight scale corresponds to the laser photoexcitation pulse. 
    }
    \label{fig:FIGs}
\end{figure}

The velocity distributions of the \ce{NH3} and \ce{ND3} beams were measured at each experimental cycle by recording the temporal profiles of the gas density using the two FGs located beyond the reaction region (see Fig. \ref{fig:setup}). 
The procedure we followed is illustrated by the representative data set displayed in Fig. \ref{fig:FIGs}, which shows the time-of-flight distributions for a seeded expansion of \ce{NH3} in He (5:95) recorded at the first (red) and second (blue) FG. Both distributions have the same overall shape and are wider (full width at half maximum (FWHM) of $\sim70~\mu$s) than the 20-$\mu$s-long valve-opening time. Consequently, the observed time-of-flight
distributions can be decomposed into multiple pairs of corresponding temporal bins, indicated by the vertical lines. One such pair, representing the part of the beam that overlaps with the $\ce{D2}(n)$ cloud in the middle of the reaction region, is shaded in black.
The longitudinal velocities $v_{\text{n},z}$ associated with the different bin pairs can be directly obtained by dividing the distance $d_{\text{FG}}$ between the two FGs by the time-of-flight difference between the bins of each pair.
The velocities given in Fig. \ref{fig:FIGs} were obtained using a $d_{\text{FG}}$ value of \SI{20.2}{cm}, which was determined in a separate calibration measurement using a pure He beam.
These velocities illustrate that the short valve-opening times and the long flight distance (see Fig. \ref{fig:setup}) enable a high degree of velocity selection (around \SI{5}{m/s} in the present case), as already pointed out in earlier studies \cite{shagam13a,jankunas14b}. 

Our measurements yielded a mean velocity $\overline{v_{\text{n},z}}=\SI{1735\pm 3}{m/s}$ for the selected bin of the ammonia:He mixtures.
Similar measurements for expansions of pure ammonia revealed much broader time-of-flight distributions, with Gaussian FWHM of up to 400~$\mu$s, indicating
 much broader distributions of velocities than for the seeded beams (see Fig. \ref{fig:FIGs}).
Based on earlier works reporting similar observations \cite{menzel_scattering_1996,gang_time_1995,torres_selected_1999}, we attribute this behavior to the formation of clusters in the expansion and the resulting heating of the expanding gas through the release of $\sim \SI{170}{meV}$ of energy per clustering ammonia molecule \cite{bobbert_fragmentation_2002}. 
In such expansions, the fastest molecules at the front of the gas pulse are primarily monomers. 
To avoid clusters in our measurements involving pure ammonia expansions, we selected velocity bins around $\overline{v_{\text{n},z}}=$\SI{1380\pm 3}{m/s} for \ce{NH3} and $\overline{v_{\text{n},z}}=$\SI{1330\pm 3}{m/s} for \ce{ND3}, which are faster than the mean beam velocities of \SI{1080}{m/s} and \SI{1050}{m/s}, respectively.

The shot-to-shot characterization of the velocity distribution of the ammonia beam with the two FGs is a crucial element of our procedure.
Slow drifts of the gas-expansion conditions, resulting from temperature variations of the valve, are automatically accounted for in the analysis.
Moreover, the known distance from the FGs to the middle of the reaction zone enables us to accurately set the longitudinal velocity $\overline{v_{\text{n},z}}$ of the ammonia molecules that overlap with the $\ce{D2}(n)$ cloud by adjusting the ammonia valve-opening trigger time. 
Finally, the high degree of transverse-velocity selection through the skimmers and the two pairs of razor blades 
implies that (i) $\overline{v_{\text{n}}}\approx \overline{v_{\text{n},z}}$ and (ii) the distribution of collision energies is entirely determined by the much broader velocity distribution of the $\ce{D2}(n)$ beam (see below).
Consequently, for each collision, the relative velocity is given by the velocity of the $\ce{D2}(n)$ molecule:
\begin{equation}\label{eq:vrel2}
|v_{\text{rel}}|\approx\sqrt{v_{\text{Ry},x}^2+v_{\text{Ry},y}^2+(v_{\text{Ry},z}-\overline{v_{\text{n}}})^2}.
\end{equation}

\subsubsection{Characterization of the $\ce{D2}(n)$ beam}\label{sec:Ryd}
To determine the transverse and the longitudinal velocity distributions for each selected $\ce{D2}(n)$-beam mean velocity $v_{\text{f}}$, we combine images of the $\ce{D2}(n)$-molecule cloud recorded using the on-axis movable MCP and particle-trajectory simulations of the trapping and deflection procedure.
The images are recorded in separate experiments carried out just before or after recording the corresponding product-ion time-of-flight spectra.
A two-dimensional Gaussian function is used to fit the images, with different widths for the $x$ and $y$ directions. 
The mean velocities along $x$ and $y$ are determined by comparing the width of the $\ce{D2}(n)$-molecule cloud at the end of the deflector (obtained in the numerical particle-trajectories simulations) and the width at the position of the MCP extracted from the images.
The mean transverse velocities ($\overline{v_{\text{Ry},x}}$,$\overline{v_{\text{Ry},y}}$) are always found to be between 10 and \SI{30}{m/s}, with $\overline{v_{\text{Ry},y}}$ typically twice as large as $\overline{v_{\text{Ry},x}}$.
In these imaging experiments, we also 
verify that the size of the $\ce{D2}(n)$-molecule cloud does not exceed the size of the ammonia beam because this would lead to an undesired loss of product-ion signal. 
For $v_{\text{f}}$-values below \SI{1100}{m/s}, we observe a significant increase of the $\ce{D2}(n)$-molecule cloud size,
and the corresponding data is not included in the analysis.

The velocity distribution of the $\ce{D2}(n)$-molecule cloud along the $z$ direction is determined using (i) the measured time-of-flight distribution to the on-axis MCP, (ii) the time at which the $\ce{D2}(n)$-molecules cloud is in the center of the reaction chamber, as determined by pulsed-field ionization, and (iii) the known distance between the center of the chamber and the on-axis MCP (see Fig. \ref{fig:setup}). 
The time-of-flight traces are very well reproduced by Gaussian distributions. Their FWHM are around \SI{20}{m/s} and depend only slightly on $v_{\text{f}}$.  
The mean values of these distributions $\overline{v_\text{Ry},z}$ slightly deviate from the $v_{\text{f}}$ values programmed on the deflector, because of imperfect acceleration over the chip.
The deviations increase with the magnitude of the acceleration.

\begin{figure}
    \centering
    \includegraphics[width=\linewidth]{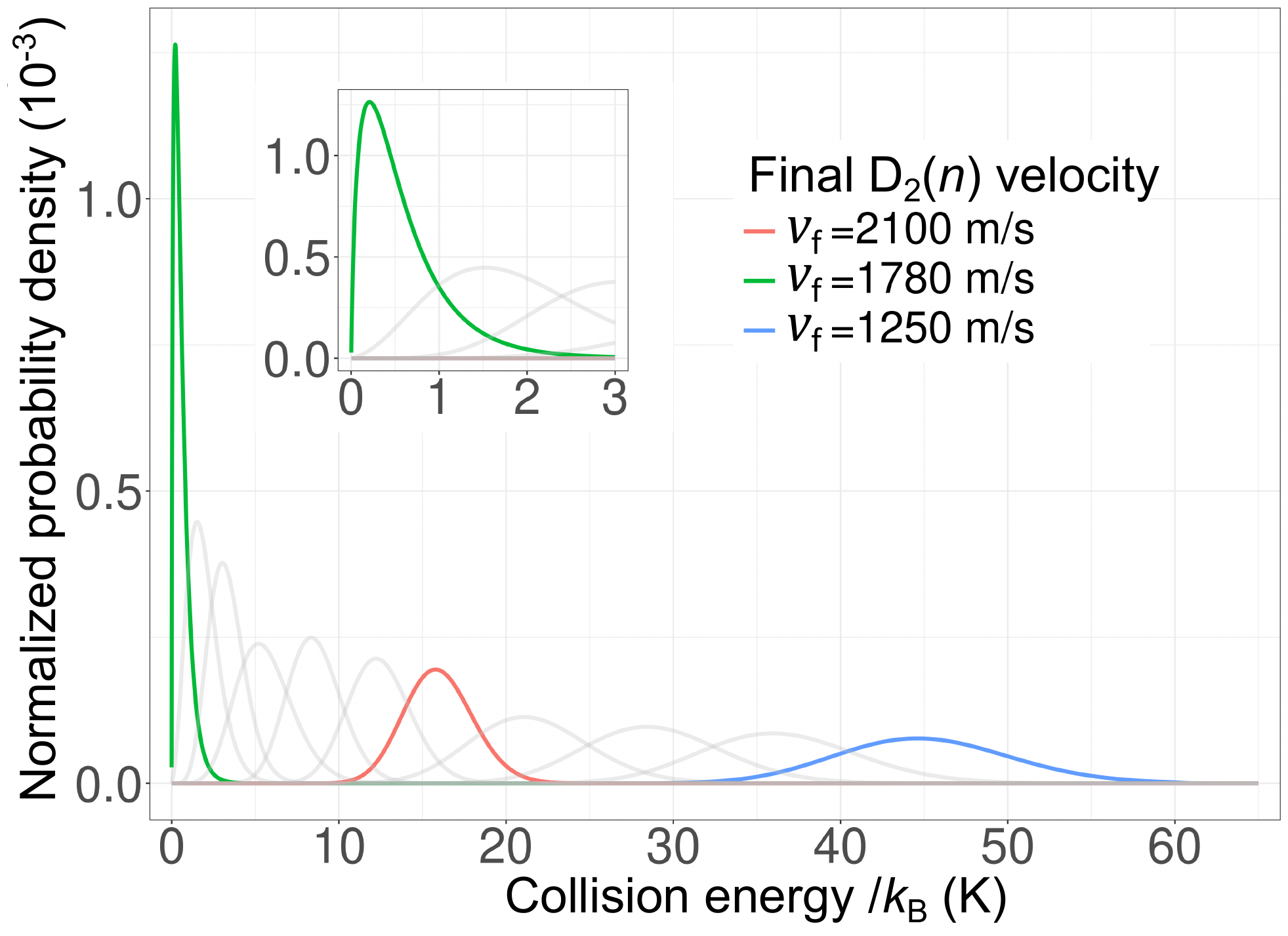}
    \caption{Total-collision-energy probability density $\rho(E_{\text{coll}};v_{\text{f}})$ determined for the $\ce{D2}(n)$ + \ce{NH3} reaction after merging a beam of \ce{NH3} seeded in He (5:95) with mean velocity $\overline{v_{\text{n},z}}=\SI{1735\pm 3}{m/s}$ with a beam of $\ce{D2}(n)$ molecules with mean longitudinal velocities between \SI{1250}{m/s} and \SI{2100}{m/s}.
    $\rho(E_{\text{coll}};v_{\text{f}}=2100$ m/s), $\rho(E_{\text{coll}};v_f=1780$ m/s) and $\rho(E_{\text{coll}};v_{\text{f}}=1250$ m/s) are displayed in red, green and blue, respectively. 
See text for details.}
    \label{fig:Ecoll}
\end{figure}

Fig. \ref{fig:Ecoll} depicts the distributions of collision-energy probability densities $\rho(E_{\text{coll}};v_{\text{f}})$ for different values of the mean velocity of the $\ce{D2}(n)$-molecule beam and for a mean velocity $\overline{v_{\text{n},z}}=\SI{1735}{m/s}$ of the supersonic beam of \ce{NH3} seeded in He. The inset shows the distributions at low collision energies on an enlarged scale and demonstrates that the collision-energy resolution is better than $k_\text{B}\cdot\SI{500}{mK}$ at the lowest collision energies.

\begin{figure*}
    \centering
    \includegraphics[width=0.82\textwidth]{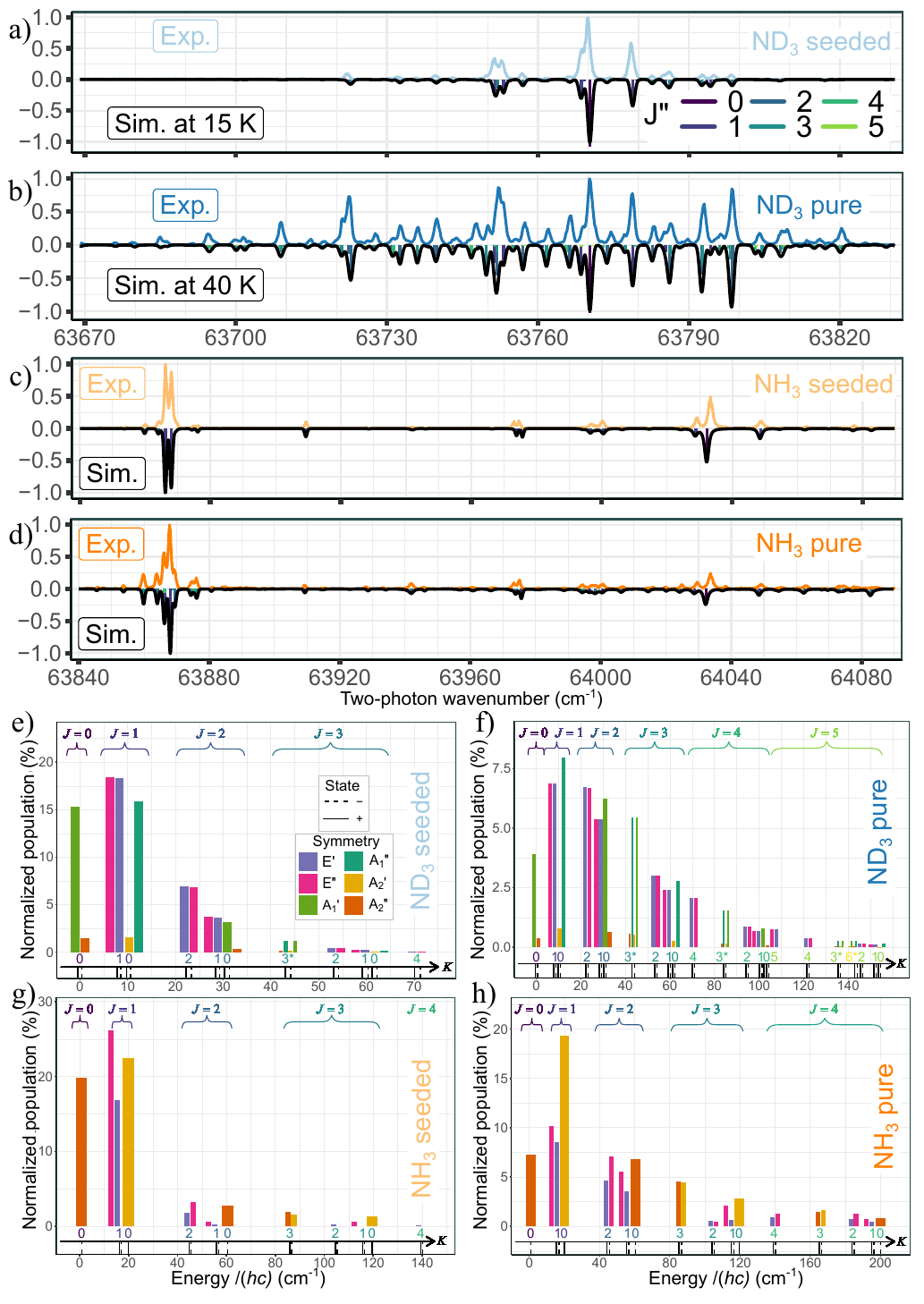}
    \caption{
    a) [b)]  Normalized (2+1) REMPI spectrum ($y$ axis in arb. units) (top: experimental; bottom, inverted: simulated) of the seeded [pure] \ce{ND3} sample, showing the transitions to the B$' (v=6)$ states. The colors of the vertical bars in the simulated spectrum indicate the rotational quantum number $J''$ of the initial state.
    c) [d)] Same as a) [b)] for \ce{NH3} for the transitions to the B$' (v=5)$ and C$' (v=0)$ states.
    e)-h) Rotational-state occupation probabilities for the seeded \ce{ND3}, pure \ce{ND3}, seeded \ce{NH3} and pure \ce{NH3} sample, respectively. The tunneling doublet is indicated as in Fig. \ref{fig:NH3andND3states}, but the splitting is multiplied by a factor of 20 for \ce{ND3}, for clarity.
    $K$ quantum numbers are indicated below the vertical bars. 
    }
    \label{fig:Trot}
\end{figure*}

\subsection{Determination of the collision-energy-dependent reaction rate coefficients}
The product ion signals $I_i(v_{\text{f}})~(i=\ce{NH3+}, \ce{NH2D+}$ for the $\ce{D2+}+\ce{NH3}$ reaction) observed at a given final velocity $v_{\text{f}}$ of the $\ce{D2}(n)$ beam
\begin{equation}\label{eq:I}
  I_i(v_{\text{f}})\propto\tau_\text{R} k_{\text{exp}}(v_{\text{f}})\left[\ce{NH3}\right]\left[\ce{D2}(n)\right]
\end{equation}
are proportional to $\tau_\text{R}$, to the densities $\left[\ce{NH3}\right]$ and $\left[\ce{D2}(n)\right]$ in the region where both beams overlap, and to an effective rate coefficient
\begin{equation}\label{eq:kexp}
k_{\text{exp}}(v_{\text{f}})=\int_{E_{\text{coll}}}k_{\text{th}}(E_{\text{coll}})\rho(E_{\text{coll}};v_{\text{f}})dE_{\text{coll}}, 
\end{equation}
where $\rho(E_{\text{coll}};v_{\text{f}})$ is determined from experimental data as explained in Section \ref{sec:Ryd}, and $E_{\text{coll}}$ is given by Eqs. \eqref{eq:Ecoll} and \eqref{eq:vrel2}.
To remove the influence of $\left[\ce{NH3}\right]$ and $\left[\ce{D2}(n)\right]$, we normalize the measured signal for each $v_{\text{f}}$ value with quantities proportional to $\left[\ce{NH3}\right]$ and $\left[\ce{D2}(n)\right]$. In the case of $\left[\ce{NH3}\right]$ (and $\left[\ce{ND3}\right]$), we use the time-of-flight profiles measured at the FGs (see black shaded areas in Fig. \ref{fig:FIGs}). In the case of $\left[\ce{D2}(n)\right]$ we use the strength of the \ce{D2+}-ion signal generated by the pulsed-field ionization when the ion-extraction pulse is applied. 
The amplitude of this pulse ($\approx \SI{30}{V/cm}$) is not large enough to efficiently field ionize the initially prepared $n=29$ Rydberg states (their field-ionization threshold is \SI{454}{V/cm}).
Consequently, the detected \ce{D2+} ion signal originates from a small fraction of the $\ce{D2}(n)$ molecules that have undergone transitions to higher Rydberg states by absorption of thermal radiation or collisions during their flight from the decelerator to the reaction region. 
When determining $k_\text{exp}$ from the measured product signal, one must correct for the fact that the \ce{D2+} signal increases with increasing $\ce{D2}(n)$ flight time and thus decreases with increasing $v_{\text{f}}$ values. 
For instance, the \ce{D2+} signal obtained at the largest $v_{\text{f}}$ value (\SI{2000}{m/s}) is 30\,\% smaller than that obtained for the slowest $v_{\text{f}}$ value (\SI{1250}{m/s}) although the number of $\ce{D2}(n)$ molecules is the same in both cases.

\subsection{Determination of the rotational temperature of the ammonia sample via (2+1) REMPI spectroscopy}\label{sec:REMPI}
The rotational temperatures of the \ce{NH3} and \ce{ND3} samples are determined by (2+1) REMPI spectroscopy of selected bands of the B$'\longleftarrow$ X and C$'\longleftarrow $X electronic transitions around \SI{313}{nm} \cite{dickinson01b}.
The pulsed laser radiation is generated by frequency doubling the output of a Nd:YAG-pumped dye laser (operated with DCM dye) in a $\beta$-barium-borate crystal. The laser wavenumber is measured with a wavemeter and the laser intensity is monitored by a fast photodiode.
The laser beam crosses the merged beam at right angles in the REMPI chamber beyond the reaction region (see Fig. \ref{fig:setup}), and the \ce{NH3+} or \ce{ND3+} ions are extracted toward an MCP detector in a direction ($y$) perpendicular to the merged beam.
To obtain reliable intensity distributions, the spectra are recorded at low enough laser intensities $I$ so that the (2+1) REMPI process is not saturated, and the ionization signal is proportional to $I^3$.
The ion signal is then normalized by dividing through $I^3$.

The rotational temperature is obtained by comparing the experimental intensity distributions with intensity distributions calculated using the program \textsc{pgopher} \cite{western17a} and molecular constants from Refs. \cite{fusina_inversion-rotation_1985,chen_measurements_2006,bentley_21_2000,nolde05a}. The procedure is illustrated in Fig. \ref{fig:Trot} a) and b), where spectra for a seeded and a pure \ce{ND3} sample are compared, respectively.
In these panels, the normalized experimental intensity distributions are compared with intensity distributions calculated for rotational temperatures $T_\text{rot}$ of \SI{15}{K} and \SI{40}{K}, respectively.
Panels e) and f) depict the rotational-state occupation probabilities at these temperatures, which were used to compute $k_\text{th}$ (see Eq. \eqref{eq:kth}).

The same procedure for the \ce{NH3} sample is illustrated in Fig. \ref{fig:Trot} c), d), g), and h). 
In the case of \ce{NH3}, significant deviations from a Maxwell-Boltzmann rotational-state population distribution are observed. 
This nonthermal behavior was already observed for \ce{NH3} in Refs. \cite{glownia_mpi_1980,scotoni_opto-thermal_1989,snavely_rotational_1981}.
However, the $T_{\textrm{rot}}$ that is closest to the observed distribution in \ce{NH3} is the same as for the pure \ce{ND3} and for the seeded \ce{ND3} samples. To determine $k_\text{th}$ in this case, we use the actually observed occupation probabilities, as reported in Fig. \ref{fig:Trot} g) and h).

\section{Results and discussion}

\begin{figure}
    \centering
    \includegraphics[width=\linewidth]{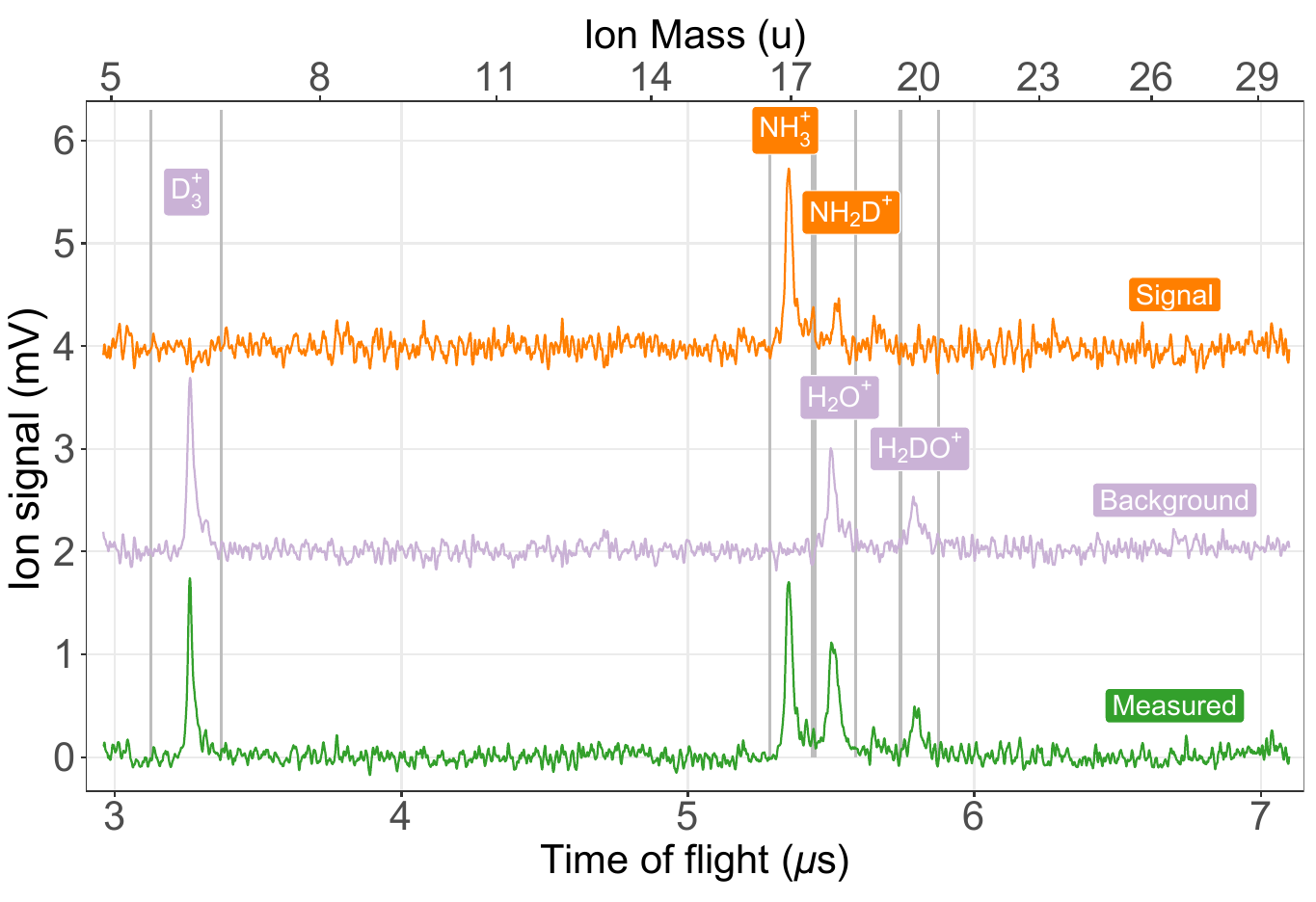}
    \caption{Experimental time-of-flight traces showing the ionic products of the reaction of $\ce{D2}(n)$ with \ce{NH3}.
    The product signals \ce{NH3+} and \ce{NH2D+} (in orange) are obtained from the green time-of-flight trace after subtraction of the background signal (purple) and integration over the temporal windows marked by the vertical lines. Offsets of \SI{2}{mV} and \SI{4}{mV} are, respectively, added to the background and signal traces for clarity. 
    }
    \label{fig:products}
\end{figure}

\begin{figure*}
    \centering
    \includegraphics[width=\linewidth]{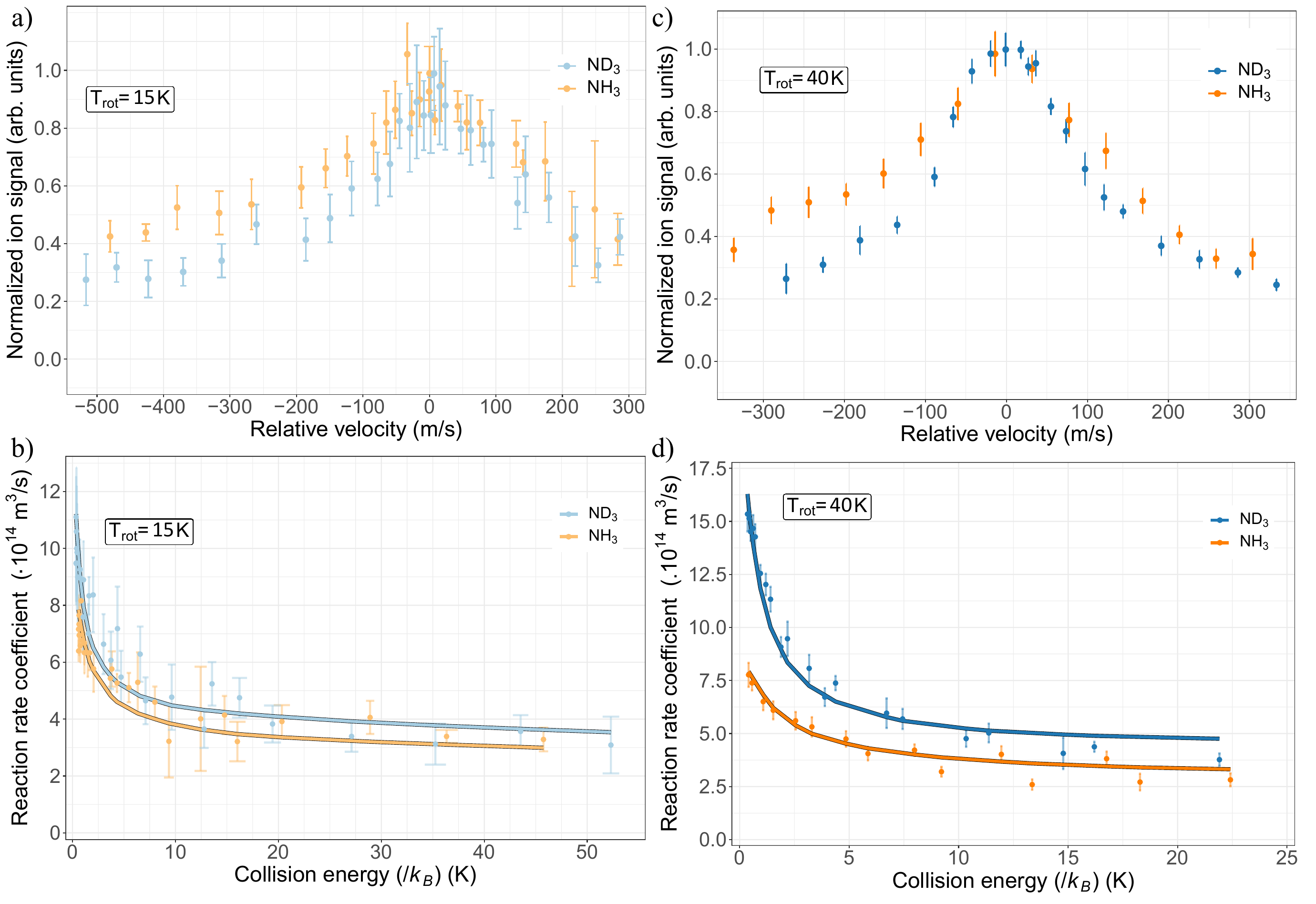}
    \caption{Product-ion signals of the $\ce{D2+}+\ce{NH3}$ and $\ce{D2+}+\ce{ND3}$ reactions as  measured (dots) for the seeded \ce{NH3} and \ce{ND3} samples (a) and for the pure \ce{NH3} and \ce{ND3} samples (c), given as a function of the nominal relative velocity. The experimental data are scaled by a global factor and compared to the calculated reaction rate coefficients averaged over the measured distributions of rotational states and collision energies $\rho(E_{\text{coll}};v_{\text{f}})$ for the seeded samples (b) and for the pure samples (d).}
    \label{fig:ratesNH3}
\end{figure*}
\subsection{Branching ratio of the $\ce{D2+}+\ce{NH3}(\ce{ND3})$ reaction }
Fig. \ref{fig:products} compares the time-of-flight spectrum obtained for the reaction $\ce{D2}(n)+\ce{NH3}$ at a collision energy of $k_\text{B}\cdot \SI{500}{mK}$ (green) with a background spectrum (light purple) recorded after delaying the \ce{NH3} pulse so that the \ce{NH3} and $\ce{D2}(n)$ gas pulses did not overlap in the reaction zone.
The latter spectrum consists of contributions from \ce{D3+}, \ce{H2O+} and \ce{H2DO+} originating from the reactions of $\ce{D2}(n)$ with \ce{D2} and \ce{H2O} molecules in the reaction-chamber background gas.
After removing these contributions by subtraction, we obtain the time-of-flight spectrum displayed in orange, which consists of a dominant \ce{NH3+} (from the charge-transfer reaction $\ce{D2+}+\ce{NH3}\longrightarrow\ce{NH3+}+\ce{D2}$) and a weaker \ce{NH2D+} signal (from the reaction $\ce{D2+}+\ce{NH3}\longrightarrow$\ce{NH2D+}$+$ H $+$ D).
The relative intensities of 5:1 of these two product channels were found not to depend on the collision energy in the range from 0 to $k_\text{B}\cdot \SI{50}{K}$ and are compatible with the earlier observations of Kim and Huntress \cite{kim75a}, who reported that 78\% (22\%) of the reactions yield \ce{NH3+}+\ce{D2} (\ce{NH2D+} + H + D). 
In the case of the $\ce{D2+}+\ce{ND3}$, the two product channels cannot be distinguished by time-of-flight mass spectrometry.

\subsection{Comparison of experimental and theoretical rate coefficients}\label{sec:comp}
Fig. \ref{fig:ratesNH3} a) and c) show how the measured product-ion signals (dots with vertical error bars representing one standard deviation), normalized by the $\ce{D2}(n)$ and ammonia densities, vary with the nominal relative velocity $\overline{v_{\text{Ry},z}}-\overline{v_{\text{n}}}$ of the $\ce{D2}(n)$ and ammonia beams, determined as explained in Section \ref{sec:methods}.
The data presented in Panel a) correspond to experiments carried out with seeded beams of \ce{NH3} and \ce{ND3} and a rotational temperature of $\sim\SI{15}{K}$. Those displayed in Panel c) were obtained with beams of pure \ce{NH3} and \ce{ND3} and a rotational temperature of $\sim\SI{40}{K}$. 
In the experiments, we do not measure absolute \ce{NH3}, \ce{ND3} and $\ce{D2}(n)$ particle densities but only relative densities. By normalizing the product-ion signals by the measured relative densities, we can nevertheless reliably determine the rate-coefficients dependence on the relative velocity and the collision-energy (see Eq. \eqref{eq:I}). 
Panels b) and d) of Fig. \ref{fig:ratesNH3} display the same data as a) and c) as a function of the nominal collision-energy $\frac{1}{2}\mu \left(\overline{v_{\text{Ry},z}}-\overline{v_{\text{n}}}\right)^2$ and multiplied by a scaling factor (one per data set) to best reproduce the calculated averaged capture rate coefficients.

The full lines in Fig. \ref{fig:ratesNH3} represent the results of the calculations of the rate coefficients taking into account the distributions of relative collision-energies and the populations of rotational levels of \ce{NH3} and \ce{ND3} determined experimentally (see Section \ref{sec:images} \& \ref{sec:REMPI}, in particular Eq. \eqref{eq:kexp}).
The calculated and experimental data in Fig. \ref{fig:ratesNH3} agree within the experimental uncertainty.
Given that no adjustable parameters were used beyond the global scaling factors for the product-ion signals, this agreement indicates that the rotationally adiabatic capture model adequately describes the collision-energy-dependent rate coefficients over the range of conditions probed in our experiments.
This agreement, in turn, makes it possible to compare experimental data acquired under different conditions and draw conclusions concerning the origin of the observed trends. 

\subsection{Influence of the collision energy}
Fig. \ref{fig:ratesNH3} reveals that in all cases, i.e., for the reactions involving both \ce{NH3} and \ce{ND3} at both rotational temperatures, the product-ion signal increases with decreasing collision energy.
This increase 
can be interpreted within the rotationally adiabatic capture model as
arising from the Stark shifts of the rotational levels of \ce{NH3} and \ce{ND3} in the field of the \ce{D2+} ion. 
The rate coefficients of states with negative Stark shifts increase much faster with decreasing collision energies than the rate coefficients of states with positive Stark shift decrease 
(see Fig. \ref{fig:theory}). This effect is general and characteristic of polar molecules. In the case of ammonia, it is enhanced by the fact that the rotational levels occur as tunneling doublets of opposite parity that are coupled by even weak electric fields.

\subsection{Influence of the rotational temperature}
Comparison of Fig. \ref{fig:ratesNH3} b) and d) enables the analysis of the effects of the increase of the rotational temperature of the ammonia samples from \SI{15}{K} to \SI{40}{K}. The main effect is an overall increase of the reaction rate coefficient 
over the full range of values of $E_{\textrm{coll}}$ investigated experimentally.
A second effect is a steeper increase of the rate coefficient at the lowest collision energies (below \SI{5}{K}) for the $T_{\text{rot}}=\SI{40}{K}$ samples.
The increase of the reaction rate coefficients with increasing rotational temperature is surprising at first sight.
One would indeed expect dipolar molecules to become less sensitive to the electric field of an ion as their rotational kinetic energy increases. Comparison of the slopes of the $(J,K,|M|,p)$=$(1,1,1,+)$, $(2,2,2,+)$ and $(3,3,3,+)$ level energies in Fig. \ref{fig:theory} d) helps to understand why this is not the case: as $J$ increases, the Stark shifts of the most high-field-seeking states, which contribute most to the product-ion signal, become larger. The corresponding rate coefficients also become larger, explaining why the rate coefficients increase with increasing rotational temperature.

Calculations of the rate coefficients (not shown) indicate a saturation of this effect beyond \SI{50}{K}, where the reaction rates become almost independent of $T_{\textrm{rot}}$, while keeping their characteristic collision-energy dependence.

\subsection{Isotope effects on the reaction rate coefficients}
Kinetic isotope effects (KIE) upon deuteration are called normal KIE if the ratio of rate coefficients of the undeuterated ($k_\text{H}$) to the deuterated samples ($k_\text{D}$) $r_{\textrm{KIE}}=\nicefrac{k_\text{H}}{k_\text{D}}>1$ and inverse KIE if $r_{\textrm{KIE}}=\nicefrac{k_\text{H}}{k_\text{D}}<1$. 
In transition-state theory, normal and inverse KIE are characteristic of ``loose" or ``tight" transition states, respectively \cite{glad_transition_1997,chen_kinetic_2017}, but such considerations do not apply for capture-limited reactions because their rate coefficients are governed by long-range interactions.
Fig. \ref{fig:ratesNH3} reveals that the reactions of \ce{D2+} with \ce{NH3} and \ce{ND3} are subject to a pronounced inverse KIE, which depends both on the collision energy, and on the rotational temperature of the ammonia samples. 
For the rotationally cold and hot samples, we observe KIEs of $r_{\textrm{KIE}}=$0.7 and $r_{\textrm{KIE}}=$0.5 at the lowest collision energies, respectively.
These large inverse KIEs originate from the different rotational and tunneling energy level structures of \ce{NH3} and \ce{ND3} (see Section \ref{sec:model}): \ce{ND3} has a higher density of states than \ce{NH3} because of its smaller rotational constant, leading to a larger fraction of the rotational population in states of large $J$ values that have large linear negative Stark shifts.
Moreover, the tunneling splitting in \ce{ND3} (\SI{0.05}{cm^{-1}}) is more than one order of magnitude smaller than that in \ce{NH3} (\SI{0.8}{cm^{-1}}).
The Stark effect in \ce{ND3} thus becomes linear at smaller electric field strength, i.e. at larger distances from the \ce{D2+} ion, leading to higher state-specific rate coefficients (compare Panels b) and e) of Fig. \ref{fig:theory}). 
These two effects make the \ce{ND3}-\ce{D2+} rotationally adiabatic interaction potentials overall more attractive than the \ce{NH3}-\ce{D2+} potentials, and explain why the capture rate coefficients of the reactions involving \ce{ND3} are larger than for \ce{NH3}.

\section{Conclusion}
In this work we have studied two reactive systems ($\ce{D2+}+\ce{NH3}$ and $\ce{D2+}+\ce{ND3}$) in the gas phase in the collision energy range from zero to $k_{\text{B}}\cdot\SI{50}{K}$ for state-selected ions, and for two different rotational temperatures $T_\text{rot}$ of the neutral molecules, measured by (2+1) REMPI spectroscopy. 
A negative dependence of the reaction rate on the collision energy was observed experimentally for both ammonia isotopologues and for both rotational temperatures. These observations could be quantitatively accounted for as arising from the charge-dipole interaction using a rotationally adiabatic capture model.

Our investigation also revealed an increase of the reaction rate coefficients with the rotational temperature, which is counter-intuitive in the classical picture of a fast rotating dipole being more difficult to orient in the electric field of an ion.
The positive effect of rotational excitation is attributed to the increase, with the rotational angular-momentum quantum number, of the high-field-seeking behavior of ammonia molecules in the field of the ion. 
The opposite influence of translational and rotational energy on the reaction rate is remarkable and in contrast with the results obtained for other ion-molecule reactions \cite{viggiano_evidence_1990, sunderlin_rotational_1994,okada_rotational_2022}.

Our study demonstrated a pronounced inverse kinetic isotope effect, the reaction of \ce{ND3} being about twice as fast as for \ce{NH3} at the lowest collision energies and $T_\text{rot}=\SI{40}{K}$.
The KIE was found to depend on both the collision and rotational energy, which might explain recent discrepancies between measurements of KIEs in ion-molecule reactions \cite{tsikritea_inverse_2021,ard_inconsistent_2022}.

\begin{acknowledgments}
We thank Josef A. Agner and H. Schmutz for their contribution to the maintenance of the experimental setup. This work was supported financially by the Swiss National Science Foundation (project number 200020B-200478).
\end{acknowledgments}

\section*{Data Availability Statement}

The data that support the findings of this study are available from the corresponding author upon reasonable request.

\end{document}